\documentstyle[prd,aps,floats]{revtex}
\begin{document}
\draft
%
%
\input epsf
\newcommand{\trh}{T_{\rm rh}}
\newcommand{\teff}{T_{\rm eff}}
\newcommand{\delcp}{\delta_{_{\rm CP}}}
\newcommand{\mh}{m_{_{\rm H}}}
\newcommand{\mw}{m_{_{\rm W}}}
\newcommand{\alphaw}{\alpha_{_{\rm W}}}
\newcommand{\ncs}{N_{_{\rm CS}}}

\renewcommand{\topfraction}{0.8}
\twocolumn[\hsize\textwidth\columnwidth\hsize\csname
@twocolumnfalse\endcsname
\preprint{IMPERIAL-TP-98/99-39, UCLA/99/TEP/7, UNIL-IPT/99-1,
hep-ph/9902449}
\title{Non-equilibrium electroweak baryogenesis at preheating
after inflation}
\author{Juan Garc{\'\i}a-Bellido}
\address{Theoretical Physics, Blackett Laboratory, Imperial College,
Prince Consort Road, London SW7 2BZ, U.K.}
\author{Dmitri Grigoriev}
\address{Institute for Nuclear Research of Russian Academy of
Sciences, Moscow 117312, Russia}
\author{Alexander Kusenko}
\address{Department of Physics, University of California, Los Angeles
CA 90095-1547, U.S.A.}
\author{Mikhail Shaposhnikov}
\address{Institute for Theoretical Physics, University of Lausanne,
CH-1015 Lausanne,
Switzerland}
\date{February 23, 1999}
\maketitle

\begin{abstract}
  
  We present a novel scenario for baryogenesis in a hybrid inflation
  model at the electroweak scale, in which the Standard Model Higgs
  field triggers the end of inflation.  One of the conditions for a 
  successful baryogenesis, the departure from thermal equilibrium, is
  naturally achieved at the stage of preheating after inflation. The
  inflaton oscillations induce large occupation numbers for
  long-wavelength configurations of the Higgs and the gauge fields, which
  leads to a large rate of sphaleron transitions. We estimate this
  rate during the first stages of reheating and evaluate the amount of
  baryons produced due to a particular type of higher-dimensional CP
  violating operator. The universe thermalizes through fermion
  interactions, at a temperature below critical, $\trh \lesssim 100$
  GeV, preventing the wash-out of the produced baryon asymmetry.
  Numerical simulations in (1+1) dimensions support our theoretical
  analyses.
\end{abstract}

\pacs{PACS number: 98.80.Cq \\
Preprint \ IMPERIAL-TP-98/99-39, UCLA/99/TEP/7, UNIL-IPT/99-1, 
hep-ph/9902449}

\vskip2pc]

\renewcommand{\thefootnote}{\arabic{footnote}}
\setcounter{footnote}{0}

\section{Introduction}

One of the most appealing explanations for the baryon asymmetry of the
universe utilizes the non-perturbative baryon-number-violating
sphaleron interactions present in the electroweak model at high
temperatures~\cite{krs,rs}. In addition to B, C and CP violating
processes, a departure from thermal equilibrium is necessary for
baryogenesis~\cite{sakharov}. The usual scenario invokes a strongly
first-order phase transition to drive the primordial plasma out of
equilibrium and set the stage for baryogenesis~\cite{rs}.  This
scenario presupposes that the universe was in thermal equilibrium
before and after the electroweak phase transition, and far from it
during the phase transition. Although there is a mounting evidence in
support of the standard Big-Bang theory up to the nucleosynthesis
temperatures, ${\cal O}(1)$ MeV, the assumption that the universe was
in thermal equilibrium at earlier times is merely a result of a
(plausible) theoretical extrapolation.  In this paper we propose a
picture of the early universe in which thermal equilibrium is
maintained only up to temperatures of the order of 100 GeV. The
earlier history of the universe is diluted by a low-scale period of
inflation, after which the universe never reheated above the
electroweak scale.

We will show that the absence of the usual thermal phase transition at
the electroweak scale does not preclude electroweak baryogenesis. In
fact, according to recent studies of reheating after inflation, the
universe could have undergone a period of ``preheating''~\cite{KLS},
during which only certain modes are highly populated, and the universe
remains very far from thermal equilibrium~\cite{non-thermal}. Such a
stage creates an ideal environment in which a substantial baryon
asymmetry could be created. The sphaleron transitions, known to cause
a baryon number violation at high temperature, may also proceed in a
system out of thermal equilibrium. In addition, the very
non-equilibrium nature of preheating may facilitate the baryon number
generation, as emphasized in Ref.~\cite{GUTB} in the context of GUT
baryogenesis.

It remains a challenge to construct a natural model with a low scale
of inflation.  The main problem is to achieve an extreme flatness of
the effective potential in the inflaton direction (i.e. the smallness
of the inflaton mass) without fine-tuning.  Although several models
have been proposed~\cite{hybrid,kt,grs}, the lack of naturalness
remains a serious problem. Perhaps recent ideas
\cite{LinKal,dvali,ADKM} related to large internal dimensions can
provide a solution.  In our paper we will not address the problem of
naturalness, but will simply assume that the electroweak-scale
inflation took place.  The main question we are going to address is
whether the electroweak baryogenesis could take place under these
circumstances. The only qualitative feature of the low-energy
inflation that is essential to us is that it produces a ``cold'' state
in which coherent bosonic fields are misplaced from their equilibrium
vacuum values.  Another mechanism that can produce a similar state is
related to strong supercooling and spinodial decomposition phase
transition which can occur, for example, in theories with radiative
symmetry breaking \cite{witten}.

As a toy model, we consider a hybrid model of inflation~\cite{hybrid},
in which the inflaton is a SU(2)$\times$U(1)-singlet and the ordinary
Higgs doublet is the triggering field that ends inflation.
Alternatively, one can view this process as one in which the inflaton
coupling to the Higgs induces dynamical electroweak symmetry breaking,
when the inflaton slow-rolls below a certain critical value. The
resonant decay of the low-energy inflaton can generate a high-density
Higgs condensate characterized by a set of narrow spectral bands in
momentum space with large occupation numbers. The system evolves
towards equilibrium while slowly populating higher and higher momentum
modes. The expansion of the universe at the electroweak scale is
negligible compared to the mass scales involved, so the energy density
is conserved, and the final reheating temperature $\trh$ is determined
by the energy stored initially in the inflaton field. We will find
model parameters such that the final thermal state has a temperature
below the electroweak scale, $\trh <T_c \sim 100$~GeV.

Sphalerons are large extended objects sensitive mainly to the infrared
part of the spectrum. We will conjecture that the rate of sphaleron
transitions at the non-equilibrium stage of preheating after inflation
can be estimated as $\Gamma_{\rm sph} \sim \alphaw^4\teff^4$, where
$\teff$ is some ``effective'' temperature associated with the long
wavelength modes of the Higgs and gauge fields that have been
populated during preheating.

Since $\trh <T_c$, the baryon-violating processes, relatively frequent
in the non-thermal condensate, are strongly suppressed as soon as the
plasma thermalizes via the interaction with fermions.  Therefore, the
baryon asymmetry created at the end of preheating is not washed out.
This is in contrast to the equilibrium electroweak baryogenesis, where
the main constraint arises from the tendency for the baryon density to
equilibrate back to zero during the slow cooling following the
electroweak phase transition. Since the energy density at the
electroweak scale is so low, the universe expansion is essentially
irrelevant and does not affect the prediction for the baryon asymmetry.

The paper is organized as follows.  In section II we discuss the hybrid
inflationary model suitable for non-equilibrium electroweak baryogenesis.
We estimate the sphaleron rates and produced baryon asymmetry in section
III.  Our estimates, based on a number of assumptions about the complicated
non-linear dynamics, are in agreement with numerical simulations discussed
in section IV.  We summarize our conclusions in section~V.

\section{An inflationary model for the electroweak baryogenesis during
preheating}

Inflation is often associated with processes occurring at the very
high energy scales, of order the Grand Unification Scale ($\sim
10^{16}$ GeV), see Ref.~\cite{book}. However, this need not be the
case. Low-scale inflation models have been
considered~\cite{hybrid,kt,grs}.  One of the side benefits of lowering
the inflation scale is avoiding the gravitino over-production
constraints. There are other particle-physics motivations for using
the TeV scale, which is associated with supersymmetry breaking in a
class of models~\cite{gauge}.  We will discuss a simple hybrid
inflation model which satisfies the constraints from cosmic microwave
background (CMB) anisotropies and large-scale structure, and at the
same time contains the desired features for a successful reheating of
the universe. We ignore completely the issue of radiative corrections,
as discussed above.

As described in the introduction, we want to construct a model with a
reheating temperature which is below that of the electroweak scale, so
that sphaleron processes are suppressed after reheating. Such a model
necessarily has a very low rate of expansion during inflation, $H\sim
\rho^{1/2}/M_P \approx 10^{-5}$ eV, which is many orders of magnitude
smaller than the mass scales we will consider. This means that
essentially all the energy density during inflation is converted into
radiation in less than a Hubble time, {\it i.e.} before the universe
has had a chance to expand significantly. This imposes a very strict
constraint on the energy scale during inflation\footnote{One can avoid
  such constraints by coupling the inflaton to some additional
  hidden-sector fields that do not contribute to the reheating of the
  observable universe. Then the potential energy density during
  inflation can be significantly larger than $\rho\sim(200\ {\rm
    GeV})^4$.}. For example, if we want the universe to reheat to
$\trh\lesssim100$ GeV, we need a model of inflation with an energy
density of order $\rho^{1/4} \sim 200$ GeV. We will construct an
example of such a model.

Hybrid inflation~\cite{hybrid} is an ingenious model of inflation, in
which the amplitude of CMB anisotropies is not necessarily related to
the GUT scale physics~\cite{book}. The idea is very simple: instead of
ending inflation via deviations from the slow-roll, it is the symmetry
breaking by a scalar field coupled to the inflaton that triggers the
end of inflation. The model can then satisfy the CMB
constraints~\cite{COBE,Bond} and allow for the electroweak-scale
inflation. We will assume that the symmetry breaking field is in fact
the Standard Model Higgs field, and that the inflaton is an additional
SU(2)$\times$U(1)-singlet scalar field.  The model thus contains two
fields, the inflaton $\sigma$ with mass $\tilde{m}$, coupled, with
coupling $g$, to the Higgs field $H^\dagger H = \phi^2/2$, with false
vacuum energy $V_0=M^4/4\lambda$ and the vacuum expectation value
$\phi_0= M/\sqrt\lambda\equiv v$,
\begin{equation}\label{pot}
V(\sigma,\phi) = {\lambda\over4}(\phi^2 - v^2)^2 +
{1\over2} \tilde{m}^2 \sigma^2 + {1\over2} g^2 \sigma^2 \phi^2 \,.
\end{equation}
During inflation, the inflaton is large, $\sigma\gg\sigma_c\equiv
M/g$, and the effective mass of $\phi$ is, therefore, large and
positive. As a consequence, the Higgs field is fixed at $\phi=0$ and
does not contribute to the metric perturbations that gave rise to the
observed CMB anisotropies. As the inflaton field slowly rolls in the
effective potential $V(\sigma)= V_0+\tilde{m}^2 \sigma^2 /2$, it will
generate the perturbations observed by COBE on large scales~\cite{COBE}. 
Eventually, the inflaton reaches $\sigma=\sigma_c$, where the Higgs
has an effective zero mass, and at this point the quantum fluctuations
of the Higgs field trigger the electroweak symmetry breaking and
inflation ends. The number of e-folds of inflation required to solve
the horizon and flatness problems is given by
\begin{equation}
\label{efolds}
N_e \simeq 34 + \ln\Big({\trh \over 100\ {\rm GeV}}\Big) \,.
\end{equation}
The fluctuations seen by COBE on the largest scales could have arisen
in this model, $N_e\simeq 34$ e-folds before the end of inflation. The
observed amplitude and tilt of CMB temperature
anisotropies~\cite{COBE,Bond}, $\delta T/T \simeq 2\times 10^{-5}$,
and $n-1\lesssim0.1$, imposes the following constraints on the model
parameters~\cite{JGBW}:
\begin{eqnarray}\label{con}
g\,\Big({v\over M_{\rm Pl}}\Big)^3{M^2\over\tilde{m}^2}
&\simeq&1.2 \times 10^{-5}\,,\\
n-1 = {1\over\pi}\,\Big({M_{\rm Pl}\over v}\Big)^2
{\tilde{m}^2\over M^2} &<& 0.1\,.
\end{eqnarray}
For example, for $v=246$ GeV (the electroweak symmetry breaking vacuum
expectation value), $\lambda \simeq 1$, and $g \simeq 0.1$, we find
$\tilde{m} \simeq 2\times10^{-12}$ eV, and it turns out that the
spectrum is essentially scale-invariant, $n-1\simeq 5\times 10^{-14}$.
These parameters give a negligible rate of expansion during inflation,
$H\simeq7\times 10^{-6}$ eV, and a reheating temperature $\trh \simeq
70$ GeV.  However, the relevant masses for us here are those in the
true vacuum, where the Higgs has a mass $\mh= \sqrt{2\lambda}\,v
\simeq 350$ GeV, and the inflaton field a mass $m = gv \simeq 25$ GeV.
Such a field, a singlet with respect to the standard model (SM) gauge
group, could be detected at future colliders because of its large
coupling to the Higgs field~\cite{hunters}.

Some comments are in order. The consideration carried out below is
qualitatively applicable also to a more complicated theory than the
minimal SM. Let us take the minimal supersymmetric standard model
(MSSM) with an additional singlet field, the inflaton $\sigma$, as an
example.  There are three SU(2) invariant couplings of the inflaton
to the Higgs doublets $H_1$ and $H_2$: $g_{11} \sigma^2
\epsilon_{\alpha\beta}H_1^{\alpha}H_1^{\beta}$, $g_{22} \sigma^2
\epsilon_{\alpha\beta}H_2^{\alpha}H_2^{\beta}$, and $g_{12} \sigma^2
\epsilon_{\alpha\beta}H_1^{\alpha}H_2^{\beta}$.  The Higgs mass
matrix of the MSSM has the eigenvalues that range from the lightest,
$\sim 100$~GeV, to the heaviest, roughly, $500$~GeV~\cite{hunters}.
In general, the inflaton-Higgs interaction is not diagonal in the
basis that diagonalizes the Higgs mass matrix in the broken-symmetry
vacuum.  In fact, the entire Higgs mass matrix is important in
determining the conditions for parametric resonance.   We will leave
the analysis of multiple Higgs degrees of freedom for future work
because it is too complicated and is not necessary to illustrate the
main idea.

\begin{figure}[t]
\centering
\hspace*{-5.5mm}
\leavevmode\epsfysize=6.5cm \epsfbox{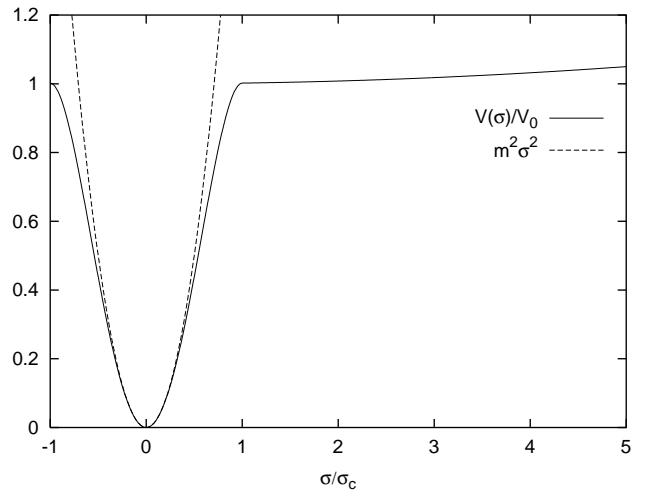}\\[3mm]
\caption[fig1]{\label{fig1} The projected effective potential
$V(\sigma)/V_0$, for the inflaton field $\sigma/\sigma_c$ after the end
of inflation. The dashed line corresponds to the $m^2\sigma^2$
approximation around the minimum of the inflaton potential. Due to the
shape of the potential at large $\sigma$, initial large-amplitude
oscillations of the field $\sigma$ are not exactly harmonic.  }
\end{figure}

\subsection{Preheating in hybrid inflation}

To study the process of parametric resonance after the end of
inflation in this model, let us recall some of the main features of
preheating in hybrid inflation~\cite{JGBL}. In hybrid models, after
the end of inflation, the two fields $\sigma$ and $\phi$ start to
oscillate around the absolute minimum of the potential, $\sigma=0$ and
$\phi=v$, with frequencies that are much greater than the rate of
expansion.  Other bosonic and fermionic fields coupled to these may be
parametrically amplified until the backreaction occurs and further
rescattering drives the system to thermal equilibrium. Initially,
rescattering of the long-wavelength modes among themselves drives them
to local thermal equilibrium, while only a very small fraction of the
short-wavelength modes are excited. The spectral density evolves
slowly towards the higher and higher momenta~\cite{TK,PR}. Eventually,
thermalization should occur through a process that breaks the
coherence of the bosonic modes, e.g. through the decay of the Higgs or
gauge fields into fermions. Such a process is very fast in the absence
of the expansion of the universe. What prevented the universe from
reheating immediately after inflation in chaotic models was the fact
that the rate of expansion in those models was much larger than the
decay rate of the inflaton, and particles did not interact with each
other until the rate of expansion dropped below the decay rate.  In
our case, the opposite is true: the rate of expansion $H\sim 10^{-5}$
eV is much smaller than the typical gauge field decay rate into
fermions, and the universe thermalizes quickly.  Since the masses are
much greater than the rate of expansion, many oscillations (of order
$10^{15}$) occur in one Hubble time~\cite{JGBL}.  It is, therefore,
possible to approximate the particle production by that in a flat
Minkowski space-time~\cite{param}.

\begin{figure}[t]
\centering
\hspace*{-5.5mm}
\leavevmode\epsfysize=6.5cm \epsfbox{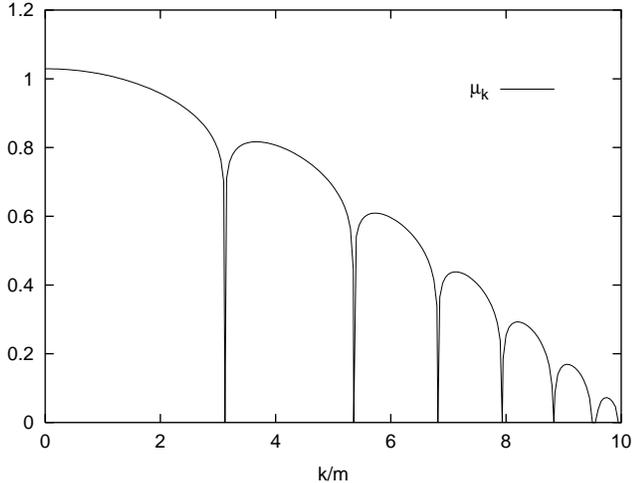}\\[3mm]
\caption[fig4]{\label{fig4} The growth parameter $\mu_k$ as a function
        of momenta $k$, in units of $m$, for a Higgs mass $\mh =
        350$~GeV.  The occupation numbers for each mode $k$
        can be obtained from $n_k=\exp(2\mu_k mt)/2$. }
\end{figure}

The evolution equation for the Fourier component of the Higgs field
that is subject to parametric resonance is approximately given by
\begin{equation}\label{Mathieu}
\ddot\phi_k + [k^2 - M^2 + 3\lambda\langle\phi^2\rangle +
g^2 \sigma^2(t)]\phi_k = 0\,.
\end{equation}
Note that this equation applies only in the case $\lambda \gg g^2$,
where we have ignored the non-linear effect of the inflaton field
$\sigma$, and in particular the cross-terms $g^2\phi_k\sigma_k$,
which do not contribute significantly before backreaction~\cite{JGBL}.
We will only use this equation for qualitative arguments, since our
quantitative results will be fully non-linear and non-perturbative,
based on numerical simulations, see Section~IV. As the inflaton
oscillates around $\sigma=0$ with amplitude $\Sigma= \sigma_c =M/g$ in
the effective potential of Fig. 1, its coupling to the Higgs will
induce
the parametric resonance with a $q$ parameter~\cite{math},
characterizing the strength of the resonance, and given by
\begin{equation}\label{qpar}
q \simeq {g^2 \Sigma^2\over4m^2}={\lambda\over4 g^2} \gg 1\,.
\end{equation}

\begin{figure}[t]
\centering
\hspace*{-5.5mm}
\leavevmode\epsfysize=6.5cm \epsfbox{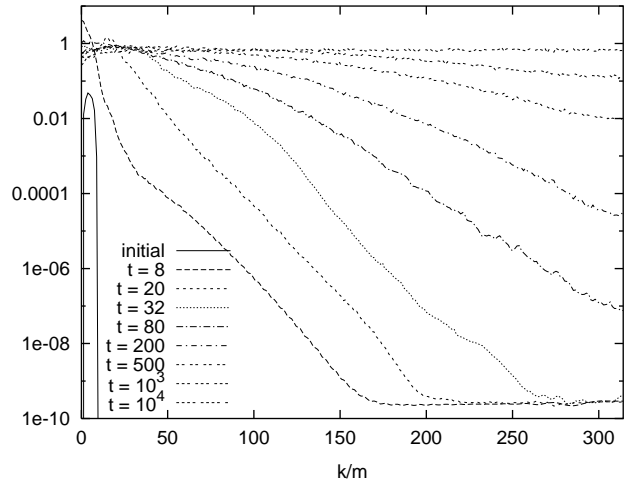}\\[3mm]
\caption[fig6]{\label{fig6} The evolution of the Higgs spectrum
  $n_k\,\omega_k$, in units of $v=246$ GeV, from time 0 to $10^4$
  $v^{-1}$, as a function of momentum, $k/m$. The initial spectrum is
  set by preheating, and contains a set of narrow bands (solid line).
  The subsequent evolution of the system leads to a redistribution of
  energy between different modes. Note how rapidly a ``thermal''
  equidistribution is reached for the long-wavelength modes. However,
  the whole Higgs spectrum approaches thermalization already in the
  middle of resonance (see Fig.~\ref{fig7} below).  }
\end{figure}

\noindent
Since we can neglect the rate of expansion, the amplitude of
oscillations $\Sigma$ does not decrease, and the resonance is
extremely long-lived. For generic values of the couplings, $g^2\sim
10^{-2}-10^{-3}$, it is, in fact, a broad resonance, $q\gg1$. Higgs
particle production occurs at the instants when $\sigma(t)=0$, and
continues until backreaction becomes important, either for the
inflaton oscillations ($\langle\phi^2\rangle \simeq
m^2/g^2$)~\cite{TK,PR,KLS} or for the effective Higgs mass
($3\lambda\langle\phi^2\rangle \simeq M^2-k_*^2+4m^2\sqrt
q$)~\cite{KLS}. Which of the two effects back-reacts first depends on
the coupling $g$.  Here $k_*=\sqrt2\,m\,q^{1/4}$ is the typical
momentum of the resonance band. For $g>0.08$ backreaction on the
inflaton mass occurs before the $\lambda$-term in (\ref{Mathieu}) is
relevant.  We have chosen $g=0.1$ for definiteness, and computed the
power spectrum of the Higgs field. For a smaller coupling $g$, the
resonance spectrum would be different, but the qualitative behavior
would be similar. In fact, it does not matter how many bands the
parametric resonance populates because after rescattering all those
bands smooth out and reach ``thermalization'' over a {\em finite}
region in momentum space~\cite{TK,PR}. In Fig.~\ref{fig4} we show the
growth parameter $\mu_k$ as a function of $k$. The typical momentum
contributing to the power spectrum, $k^2|\phi_k|^2$, is
\begin{equation}\label{kstar}
k\sim k_*/2 = m\,q^{1/4}/\sqrt2 \approx 2m \,,
\end{equation}
where the growth factor has a large value, $\mu_{\rm max} \simeq 0.9$.
This unusually large number is due to the fact that what drives the
Higgs production in this model is not the usual parametric resonance
from oscillations around the minimum of the potential~\cite{JGBL}, but
the spinodal instability responsible for the breaking of the
electroweak symmetry. In the language of Mathieu equations~\cite{math},
this corresponds to a large and negative $A=(k^2-M^2)/4m^2$ parameter,
which induces large growth factors $\mu$. On the other hand, the
occupation number of a given mode is determined from $\mu_k$ as $n_k
\simeq {1\over2}\exp(2\mu_k m t)$. This means that, within a few
oscillations, the Higgs field reaches a huge occupation number over a
range of narrow bands in momentum space.  Therefore, the Higgs
fluctuations grow exponentially with time,
\begin{equation}\label{phi2}
\langle\phi^2\rangle = {1\over2\pi^2} \int dk\,k^2 {n_k\over\omega_k}
\simeq {n_\phi(t)\over g\Sigma}\propto e^{2\mu mt}\,,
\end{equation}
with $\mu=\mu_{\rm eff}\simeq0.8$. At backreaction, the Higgs
expectation value is just of order its vacuum expectation value (VEV),
$\langle\phi^2\rangle \lesssim m^2/g^2\simeq v^2$, but continues to
grow slightly during rescattering~\cite{TK,PR}. With our set of
parameters, this happens at times $t \sim {\cal O}(1)$ GeV$^{-1}$.

In Section IV of this paper we follow a numerical approach in (1+1)
dimensions and computed the initial state from parametric resonance and
subsequent stages like rescattering and backreaction directly through
the real time evolution of the classical equations of motion for the
bosonic modes with arbitrary $k$, with all the couplings between fields
properly taken into account. This way, we have automatically included
rescattering and thermalization in the evolution.

\begin{figure}[t]
\centering
\hspace*{-5.5mm}
\leavevmode\epsfysize=6.5cm \epsfbox{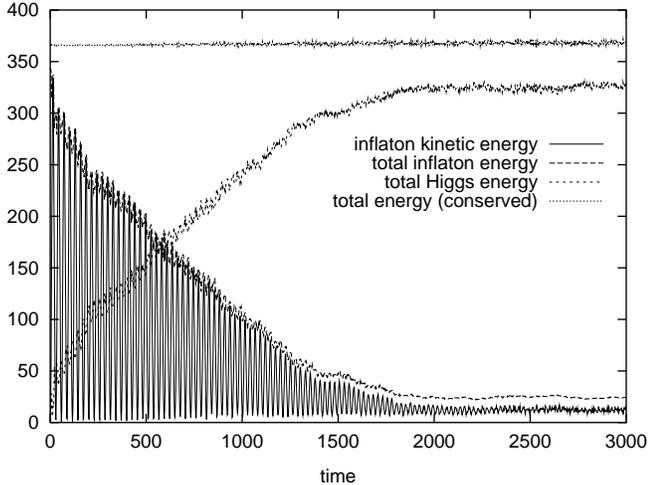}\\[3mm]
\caption[fig7]{\label{fig7} The time evolution of the inflaton energy,
the Higgs energy and the total energy. Note that the energy is measured
in units of $v$ and time in units of $v^{-1}$, see Ref.~\cite{grsnpb}.
}
\end{figure}

\subsection{Higgs coupling to W bosons}

Soon after production, Higgs particles decay predominantly into $W$
bosons with a branching ratio of order one, for $\mh=350$ GeV, and a
decay rate $\Gamma \sim 20$ GeV. One may ask whether the Higgs
oscillations may induce a resonant production of gauge bosons. It turns
out that the corresponding resonance is very narrow and insufficient
($q_{_{\rm W}}\mh = g_{_W}^2\Phi^2/4\mh \simeq 0.3\,{\rm GeV} \ll
\Gamma$, where $g_{_W}^2=4\pi\alphaw$ is the SU(2) gauge coupling and
$\Phi
\simeq v/10$ is the amplitude of the Higgs oscillations during the
first
resonance stage, see Fig.~9) for the coherent decay of the Higgs into
gauge bosons. It is therefore appropriate to use perturbation theory to
calculate the Higgs decay into the W bosons.

Since the rate of growth of the energy density of the Higgs field,
\begin{equation}\label{rho}
\rho_\phi = {1\over2\pi^2} \int dk\,k^2 n_k \omega_k
\simeq n_\phi(t)\,h\Sigma \propto e^{2\mu mt}\,,
\end{equation}
is larger than its decay rate into W bosons, i.e. $2\mu m \simeq
2\Gamma \sim 40$ GeV, we do not expect a significant depletion of the
energy density of the Higgs field during preheating, while the energy
density of the gauge bosons grows exponentially at the same rate,
$\rho_{_{\rm W}} \propto \exp(2\mu mt)$. Therefore, soon after
rescattering, most of the energy density is in the form of Higgs and
gauge fields with essentially zero momentum. It is these
long-wavelength gauge configurations that will play an important role
in inducing the sphaleron transitions, and the subsequent baryon
production.

\begin{figure}[t]
\centering
\hspace*{-5.5mm}
\leavevmode\epsfysize=6.5cm \epsfbox{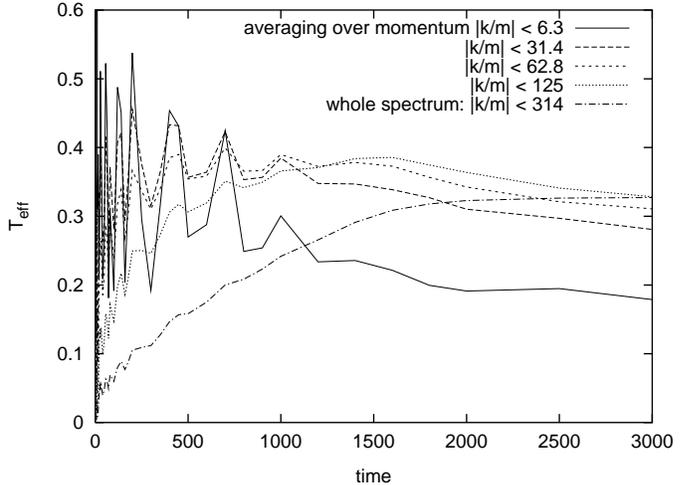}\\[3mm]
\caption[fig8]{\label{fig8} The time evolution of the effective
temperature, in units of $v$. We have averaged the Higgs power spectrum
over different low-momentum regions, and we obtain several effective
temperatures that show different time behavior.  }
\end{figure}

One of the most fascinating properties of rescattering after preheating
is that the long-wavelength part of the spectrum soon reaches some kind
of local equilibrium \cite{TK,PR}, while the energy density is drained,
through rescattering and excitations, into the higher frequency modes.
Therefore, initially the low energy modes reach ``thermalization'' at a
higher effective ``temperature'' \cite{non-thermal}, while the high
energy modes remain unpopulated, and the system is still far from true
thermal equilibrium:
\begin{equation}\label{planck}
n_k = {1\over \exp(\omega_k/T) -1} \approx {\teff\over\omega_k}
\gg 1\,.
\end{equation}
It is possible to estimate the effective ``temperature'' $\teff$ from
the conservation of energy during preheating. The energy per (long
wavelength) mode is $n_k\,\omega_k \approx \teff$, or effectively
equipartitioned. Since only the modes in the range $0<k\lesssim k_{\rm
max} \simeq 4k_*\sim 5\,m\,q^{1/4}$ are populated, we can integrate
the energy density in Higgs and gauge fields, $g_{_{\rm B}}=
1+3\times3=10$, to give, in (3+1)-dimensions,
\begin{eqnarray}
\rho_{\rm bosons} &=& g_{_{\rm
B}}\!\int\!{d^3k\over(2\pi)^3}\,n_k\,\omega_k
\simeq {g_{_{\rm B}}\over6\pi^2}\,\teff\,k_{\rm max}^3
\simeq {\lambda v^4\over4}\,, \nonumber \\
\label{teff}
{\teff\over v} &\simeq& {6\pi^2\over125}
{q^{1/4}\over g_{_{\rm B}}g} \propto g^{-1/2}\,,
\end{eqnarray}
which gives $\teff \simeq 350$ GeV.  We note that the effective
temperature depends on the value of the coupling $g$ as $\teff^4
\propto g^{-2}$, as expected~\cite{KLS}. The temperature $\teff$ is
higher than the final reheating temperature $\trh$, which is easy to
understand, since preheating is a very efficient mechanism for
populating just a few modes, into which a large fraction of the
original inflaton energy density is put. This means that a few modes
carry a large amount of energy as they come into partial equilibrium
among themselves, and thus the effective ``temperature'' is high.
However, when the system reaches a full thermal equilibrium, the same
energy is distributed between all modes, which corresponds to a much
lower temperature. In our example, thermalization of long-wavelength
modes happens at a time scale $t \sim \Gamma_{_{\rm W}}^{-1} \sim
{\cal O}(1)$~GeV$^{-1}$, where $\Gamma_{_{\rm W}}\simeq2$~GeV is the
width of the vector boson.

\begin{figure}[t]
\centering
\hspace*{-5.5mm}
\leavevmode\epsfysize=6.5cm \epsfbox{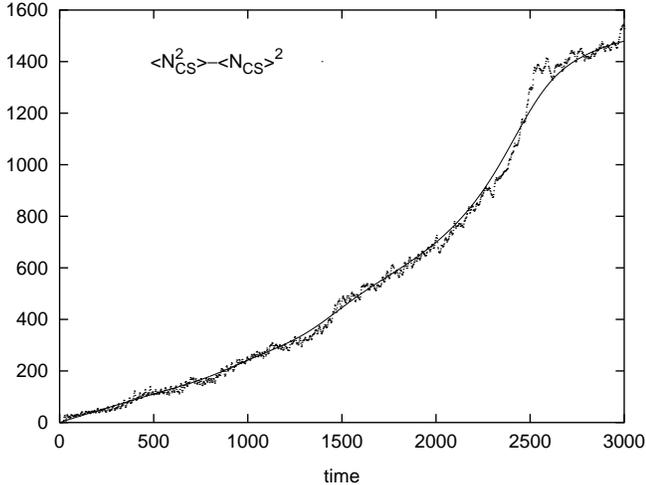}\\[3mm]
\caption[fig9]{\label{fig9} The variance of the Chern-Simons number,
i.e. $\langle\ncs^2\rangle -$ $\langle\ncs(t)\rangle^2$, as a function
of
time. The solid line is a result of smoothing out the measured values
(dots).}
\end{figure}

This is the main reason why the out of equilibrium mechanism of
preheating is so efficient in producing sphaleron transitions, since
the rate of these transitions is greatly enhanced by the higher
effective temperature. An alternative way of seeing this is by analogy
with a diffusing plasma. The rescattering of Higgs and W bosons
after preheating produces a diffusion which enhances over-the-barrier
sphaleron transitions. It is fortunate that the description of
this diffusion mechanism can be done with the use of an effective
temperature, for which the rate of sphaleron transitions can be
estimated analytically, by ignoring the higher momentum modes and the
integration over hard thermal loops~\cite{Bodeker}.

\begin{figure}[t]
\centering
\hspace*{-5.5mm}
\leavevmode\epsfysize=6.5cm \epsfbox{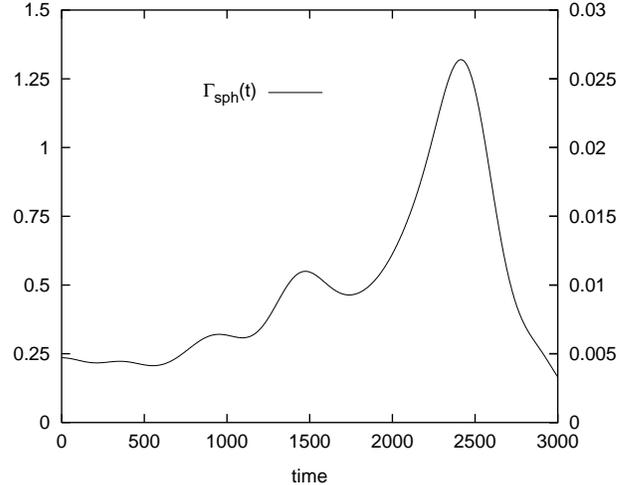}\\[3mm]
\caption[fig10]{\label{fig10} The sphaleron transition rate
$\Gamma_{\rm
sph}(t)$ is proportional to the time derivative of the Chern-Simons
variance (solid line on Fig.~\ref{fig9}). The left scale measures the
transitions per unit time per whole sample volume. The right scale
measures the sphaleron transitions per unit time per unit volume. Note
that because of high final equilibrium temperature the transitions
aren't vanishing after the resonance.  }
\end{figure}

\section{Baryon asymmetry of the universe}

It is well known that sphaleron transitions are mainly sensitive to the
long-wavelength modes in a plasma. This is because the sphaleron size,
$(\alphaw\teff)^{-1}$, is much larger than the typical Compton
wavelengths of particles in the plasma, $(2k_*)^{-1}\simeq (5m)^{-1}$.
A
simple argument then suggests that the rate of sphaleron transitions
per
unit time per unit volume should be of the order of the fourth power of
the magnetic screening length in the
plasma~\cite{armcl,khsh}.\footnote{In the symmetric phase of the
electroweak theory and in thermal equilibrium at a temperature $T$, the
typical momentum scale of sphaleron processes is $\alphaw T$
($\alphaw\simeq1/29$ is the weak gauge coupling) which is much smaller
than the average momentum of the particles in the plasma, $k\sim T$. It
was argued in Refs.~\cite{Arnold,Bodeker} that the higher momentum
modes
with typical scale greater than $g_{_W}T$ should slow down the
sphaleron
processes by an extra factor $\alphaw \log (1/\alphaw)$. During the
first stages of reheating those high frequency modes are not populated
and therefore should not be considered in our estimate.}

We, therefore, conjecture that the sphaleron transition rate during
rescattering after preheating, $\Gamma_{\rm sph}$, can be approximated
by that of a system in thermal equilibrium at some temperature $\teff$
defined in the previous section:
\begin{equation}\label{gamsph}
\Gamma_{\rm sph} \approx \alphaw^4 \teff^4\,.
\end{equation}
In the Standard Model, baryon and lepton numbers are not conserved
because of the non-perturbative processes that involve the chiral
anomaly:
\begin{equation}
\partial_\mu j_{_B}^\mu = \partial_\mu j_{_L}^\mu =
{3g_{_W}^2\over32\pi^2}\,F_{\mu\nu}\tilde F^{\mu\nu}\,.
\end{equation}
Furthermore, the sphaleron configurations connect vacua with different
Chern-Simons numbers, $N_{_{CS}}$, and induce the corresponding changes
in the baryon and lepton number, $\Delta B = \Delta L = 3 \Delta \ncs$.

\begin{figure}[t]
\centering
\hspace*{-5.5mm}
\leavevmode\epsfysize=6.5cm \epsfbox{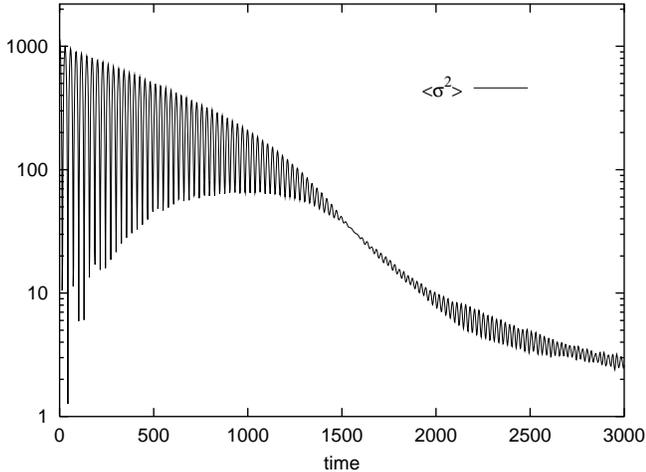}\\[3mm]
\caption[fig11a]{\label{fig11a} The expectation value of
$\langle\sigma^2\rangle$ in units of $v^2$, as a function of time. }
\end{figure}

A baryon asymmetry can be generated by sphaleron transitions in the
presence of C and CP violation. There are several possible sources of
CP
violation at the electroweak scale.  The only one confirmed
experimentally
is due to Cabibbo-Kobayashi-Maskawa mixing of quarks that introduces
some
violation of CP, but it is probably too small to cause a sufficient
baryon
asymmetry. Various extensions of the Standard Model contain additional
scalars (e.g. extra Higgs doublets, squarks, sleptons, {\it etc.}) with
irremovable complex phases that lead to C and CP violation.

We are going to model the effects of CP violation in the effective
field
theory approach.  Namely, we assume that, after all degrees of freedom
except the gauge fields, the Higgs, and the inflaton are integrated
out,
the effective Lagrangian contains some non-renormalizable operators
that
break CP. The lowest, dimension-six operator of this sort is~\cite{ms}
\begin{equation}\label{cpnonc}
{\cal O} = {\delcp\over M_{\rm new}^2}\phi^\dagger\phi \,
{3g_{_W}^2\over32\pi^2}\,F_{\mu\nu}\tilde F^{\mu\nu} \,.
\end{equation}
The dimensionless parameter $\delcp$ is an effective measure of CP
violation, and $M_{\rm new}$ characterizes the scale at which the new
physics, responsible for this effective operator, is important. Of
course, other types of CP violating operators are possible although,
qualitatively, they lead to the same picture.

Note that the operator (\ref{cpnonc}) is CP-odd but does not violate
C. Thus, in a pure bosonic theory non-equilibrium evolution can only
produce parity-odd or CP-odd configurations, but no C asymmetry. For
example, the Chern-Simons number can be produced, as it is C-even but P
and
CP-odd. C violation, necessary for baryogenesis, comes from ordinary
gauge-fermion electroweak interactions that violate C and parity, but
conserve CP. This manifests itself in the anomaly equation that relates
baryon number (C-odd but P-even operator) to the Chern-Simons number
(C$=+1$, P$=-1$). In other words, C violation in the bosonic sector of
the
theory is not required as long as it appears in the fermionic sector,
via
the electroweak interactions.

\begin{figure}[t]
\centering
\hspace*{-5.5mm}
\leavevmode\epsfysize=6.5cm \epsfbox{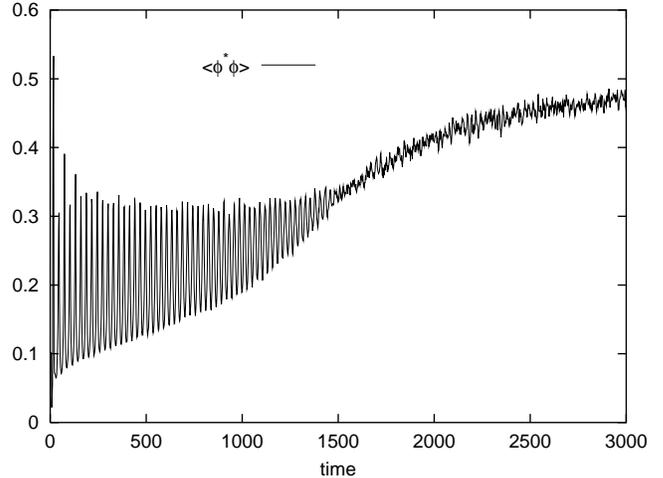}\\[3mm]
\caption[fig11b]{\label{fig11b} The expectation value of
$\langle\phi^*\phi\rangle$, in units of $v^2$, as a function of
time. The variations in $\langle\phi^*\phi\rangle$ give rise to a CP
nonconserving chemical potential $\mu_{\rm eff}$, which is the source
of
the baryon asymmetry.}
\end{figure}

If the scalar field is time-dependent, the vacua with different
Chern-Simons numbers are not degenerate. This can be
described quantitatively in terms of an effective chemical potential,
$\mu_{\rm eff}$, which introduces a bias between baryons and
antibaryons,
\begin{equation}\label{mueff}
\mu_{\rm eff}\simeq {\delcp\over M_{\rm new}^2}
{d\over dt}\langle\phi^2\rangle\,.
\end{equation}
This equation follows from Eq.~(\ref{cpnonc}) by integration by
parts.  Although the system is very far from thermal
equilibrium, we will assume that the evolution of the baryon number
$n_{_B}$ can be described by a Boltzmann-like equation, where only
the long-wavelength modes contribute,
\begin{equation}
\frac{d n_{_B}}{dt } = \Gamma_{\rm sph} {\mu_{\rm eff}\over\teff} -
\Gamma_{_B}\,n_{_B} \,,
\label{bau}
\end{equation}
where $\Gamma_{_B} = (39/2) \Gamma_{\rm sph}/\teff^3$. The temperature
$\teff$ decreases with time because of rescattering: the energy stored
in the low-frequency modes is transferred to the high-momentum modes.

\begin{figure}[t]
\centering
\hspace*{-5.5mm}
\leavevmode\epsfysize=6.5cm \epsfbox{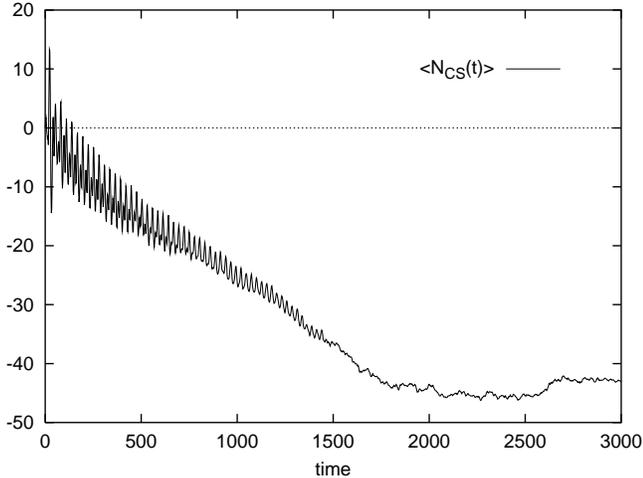}\\[3mm]
\caption[fig12]{\label{fig12}
The Chern-Simons number $\ncs(t)$ averaged over 100 runs, for
CP violating parameter $\kappa=-1$. Note that it settles at
a large and negative $\ncs$, corresponding to large baryon
production. }
\end{figure}

The rate $\Gamma_{_B}$, even at high effective temperatures, is smaller
than other typical scales in the problem. Indeed, for $\teff \sim 400$
GeV, $\Gamma_{_B} \sim 0.01$ GeV, which is small compared to the rate
of
the resonant growth of the Higgs condensate ($2\mu m\sim 40$ GeV). It
is also much smaller than the decay rate of the Higgs into W's and the
rate of W decays into light fermions.  Therefore, the last term in
Eq.~(\ref{bau}) never dominates during preheating and the final baryon
asymmetry can be obtained by integrating
\begin{equation}
n_{_B} = \int dt\,\Gamma_{\rm sph}(t) {\mu_{\rm eff}(t)\over\teff(t)}
\simeq \Gamma_{\rm sph}{\delcp\over\teff}{\langle\phi^2\rangle
\over M_{\rm new}^2}\,,
\end{equation}
where all quantities are taken at the time of thermalization.
This corresponds to a baryon asymmetry
\begin{equation}\label{nB}
{n_{_B}\over s} \simeq {45\alphaw^4\delcp\over
2\pi^2\,g_*} {\langle\phi^2\rangle
\over M_{\rm new}^2}\,\Big({\teff\over\trh}\Big)^3\,,
\end{equation}
where $g_*=g_{_{\rm B}}+(7/8)g_{_{\rm F}}\sim 10^2$ is the number of
effective degrees of freedom that contribute to the entropy density $s$
at the electroweak scale. Taking $\langle\phi^2\rangle \simeq
v^2=(246\,{\rm GeV})^2$, the scale of new physics $M_{\rm new} \sim 1$
TeV, the coupling $\alphaw\simeq 1/29$, the temperatures $\teff \simeq
350$ GeV and $\trh \simeq 70$ GeV, we find
\begin{equation}\label{BAU}
{n_{_B}\over s} \simeq 3\times10^{-8}\,\delcp\,{v^2\over M_{\rm new}^2}
\Big({\teff\over\trh}\Big)^3 \simeq 2\times10^{-7}\,\delcp\,,
\end{equation}
consistent with observations for $\delcp \simeq 10^{-3}$, which is a
reasonable value from the point of view of particle physics beyond the
Standard Model. Therefore,  baryogenesis
at preheating can be very efficient in the presence of CP violation
that
comes from new physics at $M_{\rm new} \sim 1$ TeV.

\begin{figure}[t]
\centering
\hspace*{-5.5mm}
\leavevmode\epsfysize=6.5cm \epsfbox{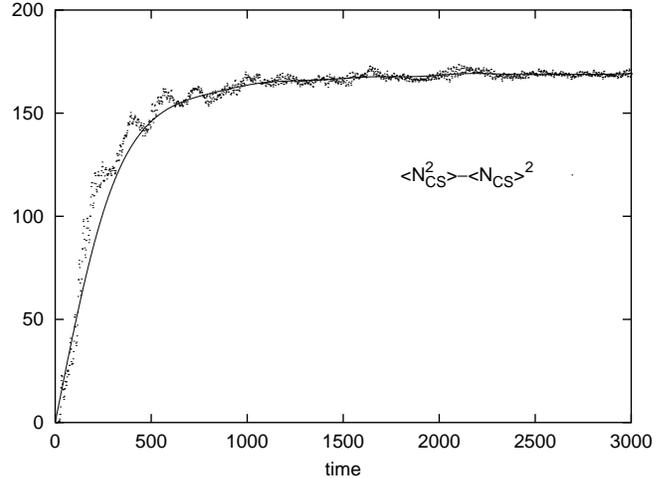}\\[3mm]
\caption[fig13]{\label{fig13} The variance of the topological
Chern-Simons number $\langle\ncs^2\rangle -$ $\langle\ncs(t)\rangle^2$
for a set of runs with total energy reduced by a factor of 4, as
compared to runs in Figs.~\ref{fig6}-\ref{fig12}.  }
\end{figure}

\section{Numerical simulations in (1+1) dimensions}

The theoretical analysis presented above was based on the conjecture
that the sphaleron transition rate can be described in terms of the
effective ``temperature'' $\teff$ as in equation (\ref{gamsph}).  This
assumption is based on the reasoning given above and seems quite
plausible.  We have also verified the validity of such description in
the (1+1)-dimensional numerical simulations.

For simplicity, we consider an Abelian Higgs model in (1+1) dimensions,
which was successfully used before for studying physics relevant to
baryogenesis~\cite{grsnpb,1p1,smit}.  The Lagrangian comprises two
scalar fields and a U(1) gauge field:

\begin{eqnarray}
{\cal L} &=& - {1\over4}F_{\mu\nu}^2 - \kappa |\phi|^2 \,
\epsilon_{\mu\nu}F^{\mu\nu} \nonumber \\
 &+& |D_\mu\phi|^2 - {\lambda}(|\phi|^2 - v^2/2)^2 \label{lagr}\\
 &+& {1\over 2} (\partial_\mu\sigma)^2 -
{1\over 2}\tilde{m}^2\sigma^2 - g^2 \sigma^2 |\phi|^2, \nonumber
\end{eqnarray}
where $D_\mu = \partial_\mu - i e A_\mu$, with $e$ the U(1) gauge
coupling, and $\epsilon_{\mu\nu}$ is the totally antisymmetric tensor
in
(1+1) dimensions. Here CP violation is induced via the $\kappa\,
\phi^*\phi\,\epsilon_{\mu\nu}F^{\mu\nu}$ term, which violates both C
and
CP. Furthermore, in (1+1) dimensions, the analogue of the chiral
anomaly is the anomalous non-conservation of the gauge invariant
fermionic current, $j_F^\mu = \bar\psi\gamma^\mu\psi$,
\begin{equation}
\partial_\mu j_F^\mu = - {e\over4\pi}\,\epsilon_{\mu\nu}F^{\mu\nu}\,,
\end{equation}
which serves as a source of B violation.

The corresponding equations of motion are:
\begin{eqnarray}
\partial_\nu F^{\mu\nu} + 2 \kappa \epsilon^{\mu\nu}
\partial_\nu |\phi|^2 &=& e j_\phi^\mu \,,\nonumber \\
D^2\phi + {2\lambda}\phi(|\phi|^2 - v^2/2) + g^2\sigma^2 \phi&=&
- \kappa \phi\,\epsilon_{\mu\nu}F^{\mu\nu} \label{firsteq} \,,\\
\partial_\mu\partial^\mu\sigma + \tilde{m}^2\sigma +
2g^2|\phi|^2 \sigma&=& 0 \,, \nonumber
\end{eqnarray}
where $j_\phi^\mu = i (\phi^*\partial^\mu \phi - \phi\partial^\mu
\phi^*) $ is the charged current of the Higgs field.

The numerical simulations track the real-time evolution of the classical
field configurations. The initial conditions are set by preheating as a set
of narrow bands in the Higgs power spectrum (Fig.~\ref{fig6}). The
real-time evolution leads to a gradual redistribution of energy between
different modes, including the inflaton field $\sigma$ itself. Note that
the Higgs field takes a very long time to reach its VEV, see
Fig.~\ref{fig11b}, due to the presence of the CP violating term in
Eq.~(\ref{firsteq}), which leads to the production of baryon number before
equilibrium. In fact, the full thermalization of the system should take a
very long time~\cite{PR,gq98}, while some other processes (e.g. interaction
with fermions from the decay of gauge fields and/or Higgs) will induce
thermalization via decoherence long before that. Thus, although it is
technically possible to reach complete thermalization of the whole system
including the inflaton, our simulations are necessarily limited to the
vector and Higgs decay time scale, of the order of $50$ inflaton
oscillations, as in Fig.~\ref{fig7}.

\subsection{The sphaleron transition rate and the effective
temperature}

As expected, the resonant inflaton decay quickly leads to a population
of the long-wavelength modes of the Higgs field, see Fig.~\ref{fig6}.
This happens after only a few oscillations of the inflaton. At this
point the long-wavelength modes contain a very large fraction of the
total energy, and that leads to a noticeable increase in $\teff$ at the
beginning of the resonance, see Fig.~\ref{fig8}.

\begin{figure}[t]
\centering
\hspace*{-5.5mm}
\leavevmode\epsfysize=6.5cm \epsfbox{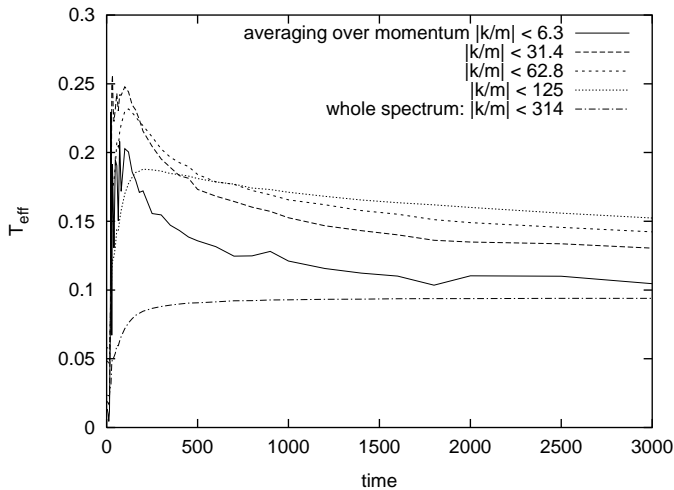}\\[3mm]
\caption[fig14]{\label{fig14}
The time evolution of effective temperature $\teff$ in units of $v$,
for the reduced-energy runs. Note the smoother rise and decline of
the effective temperature with time.}
\end{figure}

The sphaleron transitions immediately set in.  We monitor them both by
calculating the Chern-Simons number, $\ncs = \int A_1 dx^1$ in the
temporal gauge $A_0 \equiv 0$, and also by keeping track of the U(1)
winding number of Higgs field.  (Actually, no statistically significant
difference between these two quantities was observed). To get a
quantitative estimate of the transition rate we measure the variance of
$\ncs$, i.e. $\delta^2 \equiv \langle\ncs^2\rangle -
\langle\ncs(t)\rangle^2$, over an ensemble of 100 independent runs
starting from different field configurations that have the same energy
spectrum as shown in Fig.~\ref{fig6}. The preparation of the initial
configuration and other peculiarities of the numerical procedure will
be discussed in detail in a future publication.

\begin{figure}[t]
\centering
\hspace*{-5.5mm}
\leavevmode\epsfysize=6.5cm \epsfbox{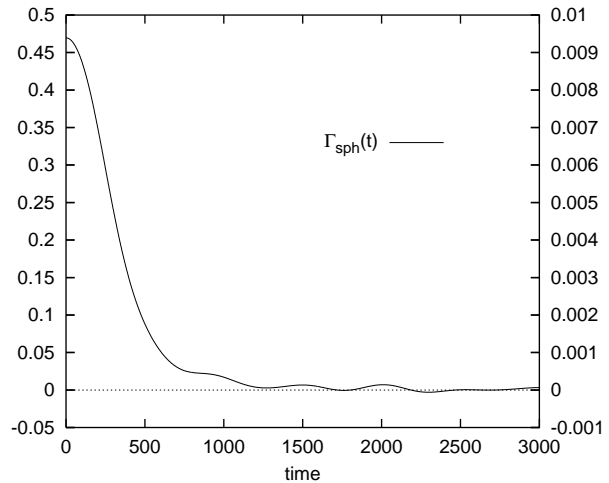}\\[3mm]
\caption[fig15]{\label{fig15} The evolution of the sphaleron rate with
time for the reduced-energy runs. The decrease of the final effective
temperature in Fig.~\ref{fig14} results in the freezing out the
equilibrium transitions soon after the resonance.  }
\end{figure}

The variance of $\ncs$ is shown in Fig.~\ref{fig9}.  Its time
derivative $d\delta^2/dt = {\rm Volume}\times\Gamma_{\rm sph}$ is
plotted in Fig.~\ref{fig10}. Note that this relation comes naturally
from the diffusion of the Chern-Simons number, $\langle\ncs^2\rangle
\simeq {\rm Volume}\times\Gamma_{\rm sph}\,t$, which follows a typical
Brownian motion~\cite{khsh}. Note that for initial parameters chosen
as in Figs.~\ref{fig6}-\ref{fig12}, the rate actually increases during
the thermalization of the Higgs field.  However, for our purposes, it
is important that we get a substantial amount of sphaleron transitions
right after the beginning of the resonance. One could slow down the
after-resonance transitions by decreasing the total energy of the
system by a factor of 4 (see Figs.~\ref{fig13}-\ref{fig17} below).
However, this decreases the net generated asymmetry considerably, due
to a decrease of the effective temperature $\teff$, see
Fig.~\ref{fig14}, and the subsequent decrease in the sphaleron rate
just after the resonance, see Fig.~\ref{fig15}. These two sets of
figures helps us gain intuition about the process of baryogenesis
during preheating in (1+1) dimensions.

\subsection{The generation of the baryon asymmetry}

As is clear from Eqs.~(\ref{lagr}) and (\ref{firsteq}), the chemical
potential $\mu_{\rm eff} \propto -\kappa\,\partial_0
\langle\phi^*\phi\rangle$ is non-zero only during the resonance.  The
energy transfer from the inflaton to the Higgs field results in a
steady
shift of $\langle\phi^*\phi\rangle$ expectation value, see
Fig.~\ref{fig11b}. This shift in VEV acts as a chemical potential and
drives the baryon asymmetry.

The baryon asymmetry generated by the non-equili\-brium sphaleron
transitions in the presence of a CP violating chemical potential
$\mu_{\rm eff}$, see Eq.~(\ref{mueff}), is observed as a non-zero
value of $\langle\ncs\rangle$ averaged over a computer-generated
ensemble. As shown in Figs.~\ref{fig12} and \ref{fig17},
$\langle\ncs\rangle$ steadily increases and eventually freezes when
the expectation value $\langle\phi^*\phi\rangle$ approaches a constant
value and the chemical potential (\ref{mueff}) vanishes.  In the early
universe, the drift of $\langle\ncs\rangle$ is eventually interrupted
by the decay of the vector and Higgs fields into
fermions.\footnote{The Higgs and vector decays into fermions are not
  included in our (1+1)-dimensional simulations. There has been recent
  progress~\cite{AS} in introducing fermions in (1+1) lattice
  simulations, but we will leave for future work such developments.}
This leads to thermalization and, as long as the reheat temperature is
sufficiently low, there is no further wash-out of the baryon
asymmetry.  We note in passing that, in our numerical simulations, the
Chern-Simons number attained at the end (i.e. the final baryon number)
is approximately linearly dependent on the CP violating parameter
$\kappa$, and, therefore, our estimate can be extrapolated to very
small values of~$\kappa$.

\begin{figure}[t]
\centering
\hspace*{-5.5mm}
\leavevmode\epsfysize=6.5cm \epsfbox{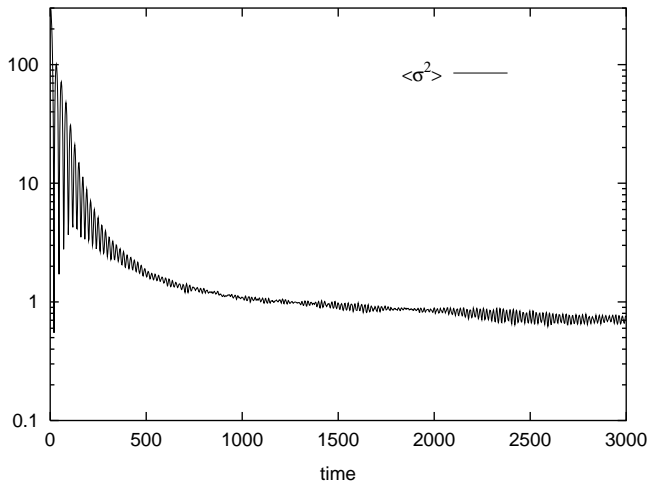}\\[3mm]
\caption[fig16]{\label{fig16} The time evolution of
$\langle\sigma^2\rangle$, in units of $v^2$,  for the
reduced-energy runs of Figs.~11-13. }
\end{figure}

\section{Conclusion}

There is no empirical evidence that a thermal electroweak phase
transition took place in the early universe.  However, since the only
well-established source of baryon number non-conservation is the
gauge sector of the Standard Model, one could argue that electroweak
baryogenesis [1] is the only explanation for the baryon asymmetry of
the universe that does not invoke any unknown B-violating new
physics.  This reasoning would favor the usual electroweak phase
transition, followed by the electroweak baryogenesis, on aesthetical
grounds.

\begin{figure}[t]
\centering
\hspace*{-5.5mm}
\leavevmode\epsfysize=6.5cm \epsfbox{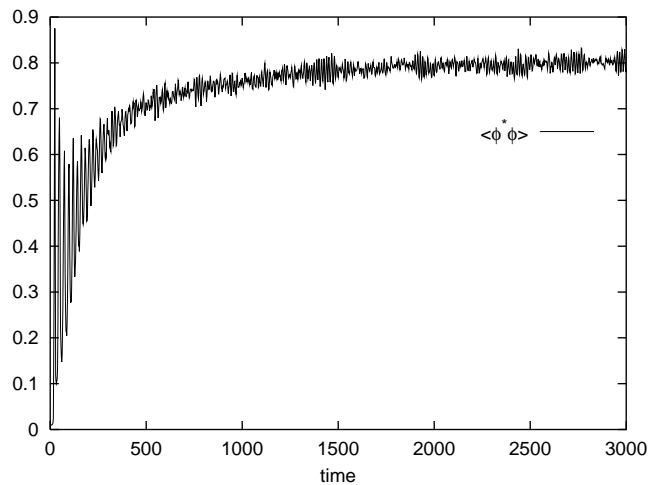}\\[3mm]
\hspace*{-5.5mm}
\leavevmode\epsfysize=6.5cm \epsfbox{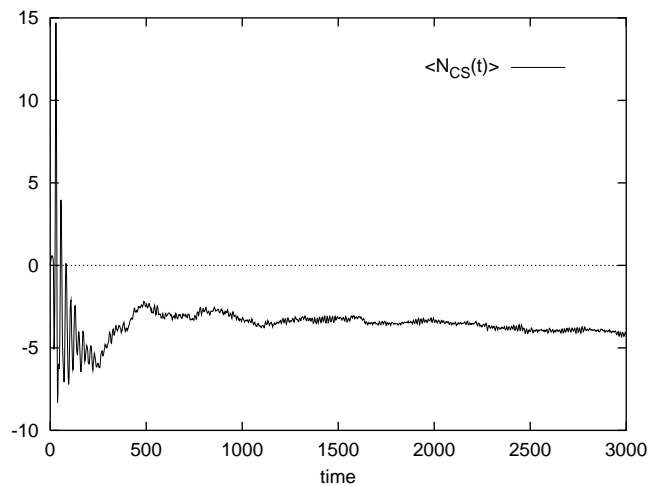}\\[3mm]
\caption[fig17]{\label{fig17} The time evolution of
$\langle\phi^*\phi\rangle$, in units of $v^2$, and
$\langle\ncs(t)\rangle$, for the reduced-energy runs
of Figs.~11-13. }
\end{figure}

In this paper we have presented an appealing alternative.  We have
shown that a new kind of electroweak baryogenesis, which still uses
only the known sources of baryon number violation, is possible even if
the reheat temperature after inflation was too low for a thermal
restoration of the SU(2)$\times$U(1) gauge symmetry.  Moreover, the
departure from thermal equilibrium, necessary for generating a
non-zero baryon number density, is naturally achieved at preheating
after an electroweak-scale inflation.  Sphaleron transitions take
place during preheating, before the thermalization of the plasma.
The baryon asymmetry can be generated through sphalerons in a manner
similar to the usual electroweak baryogenesis~\cite{rs}. When the
universe reaches thermal equilibrium, the temperature can be small
enough to suppress further baryon-violating processes, so that the
baryon asymmetry is not washed out.

\section*{Acknowledgements}

J.G.B. thanks the organizers of the ITP workshop on {\em
Non-equilibrium quantum fields}, Santa Barbara (January 1999), where
this work was presented, for a very stimulating atmosphere, and the
participants of the workshop for generous discussions. J.G.B. also
thanks Belen Gavela and Andrei Linde for enlightening comments and
suggestions. J.G.B. is supported by a Research Fellowship of the Royal
Society of London.  D.G. thanks D.V. Semikoz and M.M. Tsypin for
stimulating discussions.  D.G. is grateful to CERN TH division for
kind hospitality. D.G. work was also supported in part by RBRF grant
98-02-17493a. A.K. thanks J.M. Cornwall for helpful discussions. The
work of A.K. was supported in part by the US Department of Energy
grant DE-FG03-91ER40662. We thank Gia Dvali and Igor Tkachev for many
valuable comments.

\section*{Note added} 

After our paper was finished, we learned about a recent paper 
\cite{trodden} that also discussed baryogenesis after an
electroweak-scale inflation.


\begin{thebibliography}{99}

\bibitem{krs} V.A. Kuzmin, V.A. Rubakov, and M.E. Shaposhnikov,
  Phys. Lett. B155 (1985) 36.

\bibitem{rs} For review, see V.~A.~Rubakov and M.~E.~Shaposhnikov,
  Phys. Usp. 39 (1996) 461.

\bibitem{sakharov} A. D. Sakharov, JETP Lett. {\bf 6}, 24 (1967).

\bibitem{KLS} L.~Kofman, A.~Linde and A.~A.~Starobinsky, Phys. Rev.
Lett. {\bf 73}, 3195 (1994); Phys. Rev. D {\bf 56}, 3258 (1997).

\bibitem{non-thermal} L.~Kofman, A.~Linde and A.~A.~Starobinsky,
Phys. Rev. Lett. {\bf 76}, 1011 (1996); I.I. Tkachev, Phys. Lett.  {\bf
B376}, 35 (1996).

\bibitem{GUTB} E. W. Kolb, A.~Linde and A. Riotto, Phys. Rev. Lett.
{\bf 77}, 4290 (1996); G. W. Anderson, A.~Linde and A. Riotto,
Phys. Rev. Lett. {\bf 77}, 3716 (1996); E. W. Kolb, A. Riotto and I. I.
Tkachev, Phys. Lett. B {\bf 423}, 348 (1998).

\bibitem{hybrid} A.D. Linde, Phys. Lett. {\bf B259}, 38 (1991); Phys.
  Rev. D {\bf 49}, 748 (1994).

\bibitem{kt} L.~Knox and M.~Turner, Phys. Rev. Lett. {\bf 70} (1993)
371.

\bibitem{grs} L. Randall, M. Solja\v ci\'c, and A. H. Guth, Nucl.
  Phys. B {\bf 472}, 377 (1996); J. Garc{\'\i}a-Bellido, A. D. Linde
  and D. Wands, Phys.  Rev. D {\bf 54}, 6040 (1996).

\bibitem{LinKal} N. Kaloper and A. Linde, {\tt hep-th/9811141} (1998).

\bibitem{dvali} G.~Dvali and S.H.~Tye, {\tt hep-ph/9812483} (1998).

\bibitem{ADKM} N. Arkani-Hamed, S. Dimopoulos, N. Kaloper and
J. March-Russell, {\tt hep-ph/9903239} (1999).

\bibitem{witten} E.~Witten, Nucl. Phys. {\bf B177}, 477 (1981).

\bibitem{book} A. D. Linde {\it Particle Physics and Inflationary 
Cosmology}, Harwood Academic Press, New York, 1990.

\bibitem{gauge}  G.~Giudice and R.~Rattazzi, Phys. Rep., in press,
{\tt hep-ph/9801271} (1998).

\bibitem{COBE} C. L. Bennett et al., Astrophys. J. {\bf 464}, L1
(1996).

\bibitem{Bond} J. R. Bond, in {\em Cosmology and Large Scale
    Structure}, Les Houches Summer School Course LX, ed. R. Schaeffer
  (Elsevier Science Press, Amsterdam, 1996).

\bibitem{JGBW} J. Garc{\'\i}a-Bellido and D. Wands, Phys.  Rev. D
{\bf 54}, 7181 (1996).

\bibitem{hunters} J.~F.~Gunion, H.~E.~Haber, G.~Kane, and S.~Dawson,
{\it The Higgs Hunter's Guide}, Addison-Wesley, New York, 1996.

\bibitem{JGBL} J. Garc\'\i a-Bellido and A. D. Linde, Phys. Rev. D
        {\bf 57}, 6075 (1998).

\bibitem{TK} S.~Yu.~Khlebnikov and I.~I.~Tkachev, Phys. Rev. Lett.
{\bf 77}, 219 (1996); Phys. Rev. Lett. {\bf 79}, 1607 (1997).

\bibitem{PR} T. Prokopec and T. G. Roos, Phys. Rev. D {\bf 55},
3768 (1997).

\bibitem{param} A. A. Grib, S. G. Mamayev, and V. M. Mostepanenko, {\it
    Vacuum quantum effects in strong fields},  Friedmann Laboratory,
    St. Petersburg, 1994.

\bibitem{math} N. W. McLachlan, {\it Theory and application of Mathieu
functions}, Oxford University Press, Oxford, 1951.

\bibitem{grsnpb} D. Yu. Grigoriev, V. A. Rubakov and M. E.
Shaposhnikov,
  Phys. Lett. {\bf B216}, 172 (1989);
  Nucl. Phys. {\bf B326}, 737 (1989).

\bibitem{Bodeker} D. Bodeker, Phys. Lett. {\bf B426}, 351 (1998).

\bibitem{armcl} P. Arnold, L. McLerran, Phys. Rev. D {\bf 36}, 581
(1987).

\bibitem{khsh} S. Yu. Khlebnikov, M. E. Shaposhnikov,
Nucl. Phys. {\bf B308}, 885 (1988).

\bibitem{Arnold} P. Arnold, D. T. Son, and L. G. Yaffe, Phys. Rev. D
{\bf 55}, 6264 (1997).

\bibitem{ms} M. Shaposhnikov, Nucl. Phys. {\bf B299}, 797 (1988).

\bibitem{1p1}  D. Yu. Grigoriev, M. E. Shaposhnikov and N. G. Turok,
  Phys. Lett. {\bf B275}, 395 (1992).

\bibitem{smit} W.H. Tang and J. Smit, {\tt hep-lat/9805001} (1998).

\bibitem{gq98} D. Yu. Grigoriev, in: Procs. of the 10th Int.
Seminar QUARKS-98.

\bibitem{AS} G. Aarts and J. Smit, {\tt hep-ph/9812413} (1998).

\bibitem{trodden} L. M.~Krauss and M.~Trodden, {\tt hep-ph/9902420} (1999).

\end{thebibliography}
\end{document}